\documentstyle[12pt,cite,epsfig]{article}
\newcommand{\sect}[1]{\setcounter{equation}{0}\section{#1}}

\newcommand{\be}{\begin{equation}}
\newcommand{\ee}{\end{equation}}
\newcommand{\bea}{\begin{eqnarray}}
\newcommand{\eea}{\end{eqnarray}}

\newcommand{\p}{\partial}

\def\half{\frac{1}{2}}

\def\IB{\relax\hbox{$\inbar\kern-.3em{\rm B}$}}
\def\IC{\relax\hbox{$\inbar\kern-.3em{\rm C}$}}
\def\ID{\relax\hbox{$\inbar\kern-.3em{\rm D}$}}
\def\IE{\relax\hbox{$\inbar\kern-.3em{\rm E}$}}
\def\IF{\relax\hbox{$\inbar\kern-.3em{\rm F}$}}
\def\IG{\relax\hbox{$\inbar\kern-.3em{\rm G}$}}
\def\IGa{\relax\hbox{${\rm I}\kern-.18em\Gamma$}}
\def\IH{\relax{\rm I\kern-.18em H}}
\def\IK{\relax{\rm I\kern-.18em K}}
\def\IL{\relax{\rm I\kern-.18em L}}
\def\IP{\relax{\rm I\kern-.18em P}}
\def\IR{\relax{\rm I\kern-.18em R}}
\def\IC{\relax{\rm I\kern-.18em C}}
\def\IZ{\relax{\rm Z\kern-.5em Z}}

\begin{document}
\topmargin 0pt
\oddsidemargin 0mm

\renewcommand{\thefootnote}{\fnsymbol{footnote}}
\begin{titlepage}

\begin{flushright}
IP/BBSR/2003-09\\
hep-th/0303182\\
\end{flushright}

\begin{center}
{\Large \bf  D-branes in pp-wave spacetime with nonconstant NS-NS flux}\\
\vskip .2in
{\bf Kamal L. Panigrahi\footnote{e-mail: kamal@iopb.res.in} and
Sanjay\footnote{email: sanjay@iopb.res.in}\\
\vskip .2in
{\em Institute of Physics,\\
Bhubaneswar 751 005, INDIA}}
\end{center}
\vspace{0.1in}
\begin{center}
{\bf ABSTRACT}
\end{center}

\vskip .2in
\noindent
We find classical solutions of $D$-branes in pp-wave spacetime
with nonconstant $NS-NS$ flux. We also present $Dp-Dp^\prime$ 
bound state solutions in this background. We further analyze the 
supersymmetric properties of these brane solutions by solving 
the type IIB killing spinor equations explicitly. 

\vfill

\end{titlepage}

\sect{\bf Introduction}

Study of string theory in pp-wave background has been a subject
of much interest for theoretical physicists recently. It is known that  
pp-wave spacetime yields exact classical backgrounds
for string theory, since all curvature invariants and therefore 
all $\alpha'$ corrections, vanish\cite{klim,steif}. 
Hence pp-wave spacetimes correspond to exact conformal field theory 
backgrounds. These backgrounds are shown to be exactly solvable in 
light cone gauge\cite{met0,met1,met2}. 
Many of these are obtained from $AdS_p\times S^q$ type geometries 
in {\it Penrose} limit and are also maximally
supersymmetric\cite{penrose,blau}. 
Strings in pp-wave background are also investigated to establish 
the duality, between the supergravity
modes and the gauge theory operators in the large R-sector of 
the gauge theory\cite{malda}.  

PP-wave backgrounds with nonconstant flux are also studied
recently\cite{maoz,russo,hikida,kim,bonelli}. The worldsheet theory is 
described by nonlinear sigma model. These theories are 
the nontrivial examples of interacting theories in light cone gauge. 
The sigma model with $RR$ five form flux is supersymmetric and 
one can have linearly realized `supernumarary' supersymmetries in 
these backgrounds\cite{pope}. The corresponding sigma model with 3-form NS-NS 
and RR flux is non-supersymmetric in general\cite{russo}, unless there exists
some target space isometry and corresponding Killing vector fields,
which ensure the worldsheet supersymmetry\cite{kim}. 
Also the supernumarary supersymmetries are absent in this case due to
structure of gamma matrices. The bosonic part is same in both cases and the
two models share many properties, e.g. both are exact string theory
backgrounds and integrability structure is same etc. 
The classical solutions of $D$-branes in pp-wave background with
constant NS-NS and RR flux are already discussed in
literature\cite{kumar,sken,bain,alis,michi,kamal,rashmi,alday}.
$Dp$-branes from the worldsheet point of view are constructed 
in \cite{dabh}. Supersymmetric properties of $D$-branes in these
backgrounds are analyzed both from supergravity and worldsheet
point of view. Recently, supersymmetric properties of $D$-branes in
these backgrounds are realized more rigorously\cite{tay}.   

In earlier work, we found some supersymmetric 
solutions of $D$-branes along with its open string spectrum in
pp-wave background, arising from $AdS_3\times S^3$ geometry,
with constant three form flux\cite{kamal}. Keeping in view of the
importance of $D$-branes in understanding the nonperturbative 
as well as duality aspects of string theory, it is useful
to study them in more general backgrounds with flux being turned on. 
In this paper, we study the classical solutions of  $D$-branes 
in pp-wave spacetimes with nonconstant three form $NS-NS$ flux.  

The plan of the paper is as follows. In next section, we present
classical solutions of $Dp$ as well as $Dp-Dp^\prime$ bound states.
Section-3 is devoted to supersymmetric properties of the $D$-brane
solutions constructed in section-2. We conclude in section-4 with some 
remarks.

\sect{\bf Supergravity solutions}

In this section we present classical solutions of $Dp$ as well as
$Dp-Dp^\prime$ -branes with nonconstant $NS-NS$ three form flux transverse 
to brane worldvolume.
We start by writing down the supergravity solutions of $D$-string in
pp-wave background  with nonconstant NS-NS 
three form flux. The metric, dilaton and the field strengths are given by:   
\begin{eqnarray}
ds^2&=&f^{-{\half}}_1\left(2 dx^+dx^- + K(x_i)(dx^+)^2\right) \cr
& \cr
&+& f^{\half}_1{\sum^{8}_{m=1}}(dx^m)^2,\>\>\>
(i = 1,...,4),\cr
& \cr
H & = & \p_1 b_2(x_i)\>dx^+\wedge dx^1\wedge dx^2 + \p_3 b_4(x_i)\>dx^+\wedge
dx^3\wedge dx^4,\cr 
& \cr
e^{2\phi}&=& f_1, ~~~~F_{+-n}=\p_n f^{-1}_1,
\label{d1}
\end{eqnarray}
\noindent
with $b(x_i)$ and $K(x_i)$ satisfying the equations $\Box b
(x_i) = 0$ and 
$ \Box K (x_i) = -(\p_i b_j)^2$ respectively and 
$f_1 = 1 + {Q_1\over r^{6}}$ 
is the harmonic function in the transverse
space. We have checked that the above solution satisfies type IIB field 
equations. For constant three form flux this
solution reduces to that of ref \cite{kamal}. All other
$Dp$-brane ($p=2,...,5$) solutions can be found out by applying
$T$-duality along $x^5,...,x^8$ directions. For example:     
the classical solution for a system of $D3$-brane in such a background
is given by:
\begin{eqnarray}
ds^2&=&f^{-{\half}}_3\left(2 dx^+dx^- + K(x_i)(dx^+)^2 + (dx^{7})^2 
+ (dx^{8})^2\right) \cr
&\cr
&+& f^{\half}_3{\sum^{6}_{m=1}}(dx^m)^2,\>\>\> (i = 1,...,4),\cr
& \cr
H & = & \p_1 b_2(x_i)\>dx^+\wedge dx^1\wedge dx^2 + \p_3 b_4(x_i)\> dx^+\wedge
dx^3\wedge dx^4, \cr
& \cr
e^{2\phi}&=& 1, ~~~~F_{+-78n}=\p_n f^{-1}_3,
\label{d3}
\end{eqnarray}
\noindent
with $b(x_i)$ and $K(x_i)$ satisfying the equations $\Box b
(x_i) = 0$ and 
$ \Box K (x_i) = -(\p_i b_j)^2$ respectively and 
$f_3 = 1 + {Q_3\over r^{4}}$ is  
the harmonic function satisfying the Green function equation 
in the transverse space. 

Now we present classical solution of $D1-D5$ system as an example of
$p-p^{\prime}$ bound state in these background. The supergravity 
solution for a such system is given by:  
\begin{eqnarray}
ds^2&=&(f_1 f_5)^{-{1\over 2}}(2 dx^+dx^-
+ K(x_i) (dx^+)^2) + \Big({f_1 \over f_5}\Big)^{1\over
  2}{\sum_{m=5}^{8}}(dx^m)^2 \cr
& \cr
&+&(f_1 f_5)^{1\over 2}{\sum_{i=1}^{4}}(dx^i)^2, \cr
& \cr
e^{2\phi} &=& {f_1\over f_5} ,\cr
& \cr
H & = & \p_1 b_2(x_i)\>dx^+\wedge dx^1\wedge dx^2 + \p_3 b_4(x_i)\> dx^+\wedge
dx^3\wedge dx^4, \cr
& \cr
F_{+ - i}&=&\partial_{i}{f_1}^{-1},~~~~F_{i j k} = \epsilon_{i j k l}
\partial_{l}f_5,
\label{d1-d5}
\end{eqnarray} 
with $b(y_j)$ and $K(x_i)$ satisfying the equations $\Box b
(x_i) = 0$ and 
$ \Box K (x_i) = -(\p_i b_j)^2$ respectively and 
$f_1 = 1 + {Q_1\over
r^{2}}$ and  $f_5 = 1 + {Q_5\over
r^{2}}$ are the harmonic functions of $D1$ and $D5$-branes in common
transverse space. One can check that the above ansatz do satisfy type
IIB field equations.

\sect{Supersymmetry Analysis}

In this section we present the supersymmetry of the solutions
described earlier in section (2).
 
The supersymmetry variation of dilatino and 
gravitino fields of type IIB supergravity in ten dimension, 
in string frame, is given by \cite{schwarz,fawad}:
\begin{eqnarray}
\delta \lambda_{\pm} &=& {1\over2}(\Gamma^{\mu}\partial_{\mu}\phi \mp
{1\over 12} \Gamma^{\mu \nu \rho}H_{\mu \nu \rho})\epsilon_{\pm} + {1\over
  2}e^{\phi}(\pm \Gamma^{M}F^{(1)}_{M} + {1\over 12} \Gamma^{\mu \nu
  \rho}F^{(3)}_{\mu \nu \rho})\epsilon_{\mp},\\
\label{dilatino}
\delta {\Psi^{\pm}_{\mu}} &=& \Big[\partial_{\mu} + {1\over 4}(w_{\mu
  \hat a \hat b} \mp {1\over 2} H_{\mu \hat{a}
  \hat{b}})\Gamma^{\hat{a}\hat{b}}\Big]\epsilon_{\pm} \cr
& \cr
&+& {1\over 8}e^{\phi}\Big[\mp \Gamma^{\mu}F^{(1)}_{\mu} - {1\over 3!}
\Gamma^{\mu \nu \rho}F^{(3)}_{\mu \nu \rho} \mp {1\over 2.5!}
\Gamma^{\mu \nu \rho \alpha \beta}F^{(5)}_{\mu \nu \rho \alpha
  \beta}\Big]\Gamma_{\mu}\epsilon_{\mp},
\label{gravitino}
\end{eqnarray}
where we have used $(\mu, \nu ,\rho)$ to describe the ten
dimensional space-time indices, and hat's represent the corresponding
tangent space indices.
Solving the above two equations for D-string solution as given 
in (\ref{d1}), we get several conditions on the spinors. 
First, the dilatino variation gives:
\begin{eqnarray}
\frac{f_{1,\hat{n}}}{f_1}
\Gamma^{\hat{n}}\epsilon_{\pm}
\mp f^{\frac{1}{4}}_1(\p_{\hat{i}} b_{\hat{j}})\Gamma^{\hat{+}\hat{i}\hat{j}} 
\epsilon_{\pm} -
\frac{f_{1,\hat{n}}}{f_1}
\Gamma^{\hat{+}\hat{-}\hat{n}}\epsilon_{\mp} = 0,
\label{dilt-1}
\end{eqnarray}
Gravitino variation gives the following conditions on the spinors:
\begin{eqnarray}
\label{g+}
\delta \psi_+^{\pm} &\equiv &\left(\partial_{+}+{1\over 4}
{\p_{\hat n} (K f^{-{1\over 4}}_{1})} \Gamma^{\hat +\hat n}   
\mp{1\over4}(\p_{\hat i} b_{\hat j})(\Gamma^{\hat{i}\hat{j}})\right)
\epsilon_{\pm} \cr
& \cr
&-& {1\over 8} \Gamma^{\hat +\hat -
\hat n}{K f_{1,\hat n} \over f^{5/4}_1}\Gamma^{\hat
+}\epsilon_{\mp}=0 \\
\label{g-}
\delta \psi^{\pm}_{-} &\equiv& \partial_{-}\epsilon_{\pm}=0 \\
\label{gn}
\delta \psi_n^{\pm}&\equiv& \left(\partial_{n}
 - {1\over 8} {f_{1,n}\over f_1}\right) \epsilon_{\pm}, ~~~~
(n=5,...,8)\\
\label{gi}
\delta \psi_i^{\pm} &\equiv& \left(\partial_{i}\mp\frac{\delta_{i \hat i}}{4}f^\half_1(\p_{\hat i}b_{\hat j})
\Gamma^{\hat{+}\hat{j}} - {1\over 8} {f_{1,i}\over f_1}\right) 
\epsilon_{\pm},~~~~(i=1,...,4).
\end{eqnarray}
In writing the above equations we have used the brane supersymmetry
condition:
\begin{eqnarray}
\Gamma^{\hat +\hat -}\epsilon_{\pm} = \epsilon_{\mp}.
\label{susyb}
\end{eqnarray}
Taking derivative of the eqn. (\ref{gi}) with respect to $\p_{\hat k}$
and subtracting the derivative of $\p_{\hat k}$ equation with respect to
$\p_{\hat i}$, we get
\be
(\p_{\hat k}\p_{\hat i} b_{\hat j})\Gamma^{\hat{+}\hat{j}}\epsilon_{\pm}=0,
\ee
which can be satisfied for nonconstant $\p_{\hat i}b_{\hat j}$ only if $\Gamma^{\hat 
  +}\epsilon_{\pm} = 0$.
Using $\Gamma^{\hat 
  +}\epsilon_{\pm} = 0$. and brane supersymmetry condition
(\ref{susyb}), the dilatino variation (\ref{dilt-1}) is satisfied. 
Using  $\Gamma^{\hat 
  +}\epsilon_{\pm} = 0$, the supersymmetry conditions (\ref{gn}) 
and (\ref{gi}) are solved by spinors: $\epsilon_{\pm} = exp(-{1\over
  8} ln f_1)\epsilon^0_{\pm}$, with $\epsilon^0_{\pm}$ being a function of 
 $x^+$ only. Since $\epsilon^0_{\pm}$ is independent of $x^i$ and
 $(\p_ib_j)$ is a function of $x^i$ only, the gravitino variation
 gives the following conditions to have nontrivial solutions:
\be
(\p_{\hat i}b_{\hat j})(\Gamma^{\hat{i}\hat{j}})\epsilon^0_{\pm}=0
\label{gamma}
\ee
and
\be
\p_+\epsilon^0_\pm = 0.
\ee
The condition $\Gamma^{\hat +}\epsilon_{\pm} = 0$ breaks sixteen 
supersymmetries. The number of remaining supersymmetries depend 
on the existence of constant 
$\epsilon^0_\pm $ solutions of the equation (\ref{gamma}). For the particular
case when $H_{+12} = H_{+34}$, the equation (\ref{gamma}) gives the condition:
\be
(1 - \Gamma^{\hat 1\hat 2\hat 3\hat 4})\epsilon^0_{\pm} = 0.
\label{hat}
\ee 
Therefore in this case, the $D$-string solution
(\ref{d1}), preserves  $1/8$ supersymmetry. 
Similarly, one can show that the $D3$-brane solution 
(\ref{d3}) also preserves $1/8$ supersymmetry. 

Next, we will analyze the supersymmetry properties of  $(D1-D5)$
system that is described in eqn. (\ref{d1-d5}) of the previous section.

The dilatino variation gives the following conditions on the spinors:
\begin{eqnarray}
{f_{1,\hat i}\over f_1}
\left(\Gamma^{\hat i}\epsilon_{\pm} - \Gamma^{\hat +\hat -\hat
i}\epsilon_{\mp}\right)\mp 
(f_1 f_5)^{\frac{1}{4}}(\p_{\hat i}b_{\hat j})\Gamma^{\hat {+}\hat{i}\hat{j}}\epsilon_\pm -{f_{5,\hat i}\over f_5}
\left(\Gamma^{\hat i}\epsilon_{\pm} + {1\over 3!}\epsilon_{\hat i \hat 
    j \hat k \hat l}\Gamma^{\hat j\hat k\hat l}\epsilon_{\mp}\right)=0
\label{d1-d5-dila}
\end{eqnarray}

On the other hand, the gravitino variation gives:
\begin{eqnarray}
\delta\psi_+^{\pm}&\equiv &\partial_+ \epsilon_{\pm}+ 
{1\over 4}
{\p_{\hat n} (K (f_1 f_5)^{-{1\over 4}})} \Gamma^{\hat +\hat n}\mp {1\over
  4}
(\p_{\hat i}b_{\hat j})\Gamma^{\hat{i}\hat{j}}\epsilon_\pm \cr
&\cr
&-& {1\over 8}(f_1 f_5)^{-\frac{1}{4}}\left( \Gamma^{\hat +\hat - \hat n}{K
    f_{1,\hat n} \over f_1} + \Gamma^{\hat m_1...\hat m_3}\epsilon_{\hat
    {m_1},...,\hat{m_3},\hat n} {K f_{5,\hat n} \over f_5}\right)\Gamma^{\hat+}\epsilon_{\mp}=0 \\
\label{d1-d5-g+}
\delta\psi_-^{\pm} &\equiv& \partial_- \epsilon_{\pm} = 0,~~~~~
\delta\psi_m^{\pm} \equiv \partial_m \epsilon_{\pm} = 0,\\
\label{d1-d5-g-}
\delta\psi_i^{\pm}&\equiv &\partial_i \epsilon_{\pm} 
 - {\delta_{i\hat i}\over 4}(f_1 f_5)^{\frac{1}{2}} (\p_{\hat i}b_{\hat j})\Gamma^{\hat{+}\hat{j}}+ \frac{1}{8}\left[{f_{1,i}\over f_1} +
{f_{5,i}\over f_5}\right] \epsilon_{\pm} = 0.\>\>\>
\label{d1-d5-gi}
\end{eqnarray}
In writing down the above gravitino variations we have once again
made use of the brane conditions:
\be
\Gamma^{\hat i}\epsilon_{\pm} - \Gamma^{\hat +\hat -\hat
i}\epsilon_{\mp} = 0,
\label{d1-d5-susy}
\ee
and 
\be
\Gamma^{\hat i}\epsilon_{\pm} + {1\over 3!}\epsilon_{\hat i \hat 
    j \hat k \hat l}
\Gamma^{\hat j\hat k\hat l}\epsilon_{\mp} = 0.
\label{d1-d5-susy1}
\ee

Taking derivative of the eqn. (\ref{d1-d5-gi}) with respect to $\p_{\hat k}$
and subtracting the derivative of $\p_{\hat k}$ equation with respect to
$\p_{\hat i}$, we get
\be
(\p_{\hat k}\p_{\hat i} b_{\hat j})\Gamma^{\hat{+}\hat{j}}\epsilon_{\pm}=0,
\ee
which can be satisfied for nonconstant $\p_{\hat i}b_{\hat j}$ only if $\Gamma^{\hat 
  +}\epsilon_{\pm} = 0$.

Using $\Gamma^{\hat 
  +}\epsilon_{\pm} = 0$ and brane supersymmetry conditions
(\ref{d1-d5-susy}) and(\ref{d1-d5-susy1}) , the dilatino condition 
(\ref{d1-d5-dila}) is satisfied. Using  $\Gamma^{\hat 
 +}\epsilon_{\pm} = 0$, the supersymmetry condition
(\ref{d1-d5-gi}) is solved by spinors: $\epsilon_{\pm} = exp(-{1\over 8} ln
 (f_1f_5))\epsilon^0_{\pm}$, with $\epsilon^0_{\pm}$ being a function of 
 $x^+$ only. Since $\epsilon^0_{\pm}$ is independent of $x^i$ and
 $(\p_ib_j)$ is a function of $x^i$ only, the gravitino variation
 gives the following conditions to have nontrivial solutions:
\be
(\p_{\hat i}b_{\hat j})(\Gamma^{\hat{i}\hat{j}})\epsilon^0_{\pm}=0
\label{d1-d5-susy2}
\ee
and
\be
\p_+\epsilon^0_\pm = 0.
\ee

Once again, the number of supersymmetries depend on the existence of
solutions of equation (\ref{d1-d5-susy2}). For the particular case when 
$H_{+12} = H_{+34}$, the $D1-D5$ bound state solution (\ref{d1-d5})
also preserves $1/8$ supersymmetry.

\sect{Conclusion}

In this paper we have constructed the supergravity solutions of 
$Dp$ as well as $Dp-Dp^{\prime}$ branes in pp-wave background with 
nonconstant $NS-NS$ flux. The supersymmetric properties of these
solutions are also verified by analyzing the type IIB killing spinor
equations explicitly. All the solutions presented here
are shown to preserve $1/8$ supersymmetry and the supernumarary
supersymmetry is absent for the background presented in this
paper. $D$-brane solutions with nonconstant $RR$ flux can be found out 
by applying $S$-duality transformation on the solutions presented
here, which will be generalization of those given in \cite{kamal}. 
The $D$-brane solutions presented here have 
the interpretation of $D$-branes in nonsupersymmetric sigma model 
of \cite{russo}. It is also desirable to analyze them from the worldvolume 
point of view following the procedure of \cite{hikida}.    

\vspace{5mm}
\noindent
{\bf Acknowledgment:} We thank A. Kumar, R. R. Nayak and A. Biswas
for useful discussions. We would also like to thank S. F. Hassan and
N. Kim for comments on the earlier version of the paper.



\end{document}